\documentstyle[aas2pp4]{article}

\begin{document}

\title{Evidence from quasi-periodic oscillations for a millisecond
pulsar in the low mass x-ray binary 4U~0614+091}

\authoremail{kaaret@astro.columbia.edu}

\author{E.  Ford\altaffilmark{1}, P. Kaaret\altaffilmark{1},
M.Tavani\altaffilmark{1}, D. Barret\altaffilmark{2}, P.
Bloser\altaffilmark{2}, J. Grindlay\altaffilmark{2}, B.A.
Harmon\altaffilmark{3}, W.S.  Paciesas\altaffilmark{4,3}, and S.N.
Zhang\altaffilmark{5,3} }

\altaffiltext{1}{Department of Physics and Columbia Astrophysics
Lab, Columbia University, 538 W. 120th Street, New York, NY 10027}
\altaffiltext{2}{Harvard Smithsonian Center for Astrophysics, 60
Garden Street, Cambridge, MA 02138} \altaffiltext{3}{NASA/Marshall
Space Flight Center, ES 84, Huntsville, AL 35812}
\altaffiltext{4}{University of Alabama in Huntsville, Department
of Physics, Huntsville, AL 35899} \altaffiltext{5}{University
Space Research Association/MSFC, ES 84, Huntsville, AL 35812}

\begin{abstract}

We have detected quasi-periodic oscillations (QPOs) near 1~kHz from
the low mass X-ray binary 4U 0614+091 in observations with XTE.  The
observations span several months and sample the source over a large
range of X-ray luminosity.  In every interval QPOs are present above
400~Hz with fractional RMS amplitudes from 3 to 12\%.  At high count
rates, two high frequency QPOs are detected simultaneously.  The
difference of their frequency centroids is consistent with 323~Hz in
all observations.  During one interval a third signal is detected at
$328\pm2$~Hz.  This suggests the system has a stable `clock' which is
most likely the neutron star with spin period 3.1~msec.  Thus, our
observations and those of another neutron star system by Strohmayer et
al. (1996) provide the first evidence for millisecond pulsars within
low-mass X-ray binary systems and reveal the `missing-link' between
millisecond radiopulsars and the late stages of binary evolution in
low mass X-ray binaries (Alpar et al.  1982).  We suggest that the
kinematics of the magnetospheric beat-frequency model (Alpar and
Shaham 1985) applies to these QPOs.  In this interpretation the high
frequency signal is associated with the Keplerian frequency of the
inner accretion disk and the lower frequency `beat' signal arises from
the differential rotation frequency of the inner disk and the spinning
neutron star.  Assuming the high frequency QPO is a Keplerian orbital
frequency for the accretion disk, we find a maximum mass of $1.9
M_{\odot}$ and a maximum radius of 17~km for the neutron star.

\end{abstract}


\section{Introduction}

High frequency quasi-periodic oscillations (QPOs) have now been
discovered with XTE in at least seven low mass X-ray binaries
(LMXBs) (e.g.  van der Klis et al.  1996a, Strohmayer et al.  1996,
Berger et al. 1996, Zhang et al. 1996).  The fast variability of these 
signals is a direct result of the short dynamical time scale in the region 
near the compact object where the emission is produced.  Study of these
QPO phenomena addresses questions about the accretion flow around
the central compact object and the nature and evolution of the
compact object itself.

Here we present the discovery of high frequency QPOs in
4U~0614+091 (Ford et al. 1996a).  The x-ray source 4U~0614+091 has been 
identified as an X-ray burster (Swank et al.  1978; Brandt et al.  1992; 
Brandt 1994).  Its luminosity and color behavior make it a probable atoll 
source (Singh and Apparao 1995).  4U~0614+091 has been detected up to
100~keV with episodes of bright hard X-ray emission anticorrelated
with the soft X-ray flux (Ford et al. 1996b).

The XTE observations of 4U~0614+091 constitute the most extensive data
set of this new phenomenon to date.  We have measured the QPOs over a
wide range in frequency from 480~Hz to 1150~Hz.  Their behavior is
relatively simple, being determined mainly by the source luminosity.
The observations, analysis and results are discussed in Sections 2 and
3. In Section 4, we argue that 4U~0614+091 harbors a millisecond
pulsar, a fact which has implications for pulsar evolution
scenarios. We interpret the QPO production in terms of a simple model
and use the QPO frequency to place limits on the mass and radius of
the neutron star.

\section{Observations and Analysis}

We observed 4U~0614+091 with the Rossi X-ray Timing Explorer (Bradt et
al.  1993) for 10~ks starting UTC 4/22/96 19:18:43, 33~ks beginning
4/24/96 13:18:37, and 16~ks beginning 8/6/96 20:05:00.  The following
results utilize Proportional Counter Array (PCA) (Zhang et al.  1993)
data with 122~$\mu$s time resolution and good sensitivity from
approximately 1 to 30 keV.  The observations of 4U 0614+091 divide
into intervals of continuous coverage with typical durations of 3000 s
separated by earth occultations and/or SAA passages.

Power spectra are generated from count rate data binned in 122~$\mu$s
intervals in consecutive windows of 1~s duration; yielding a Nyquist
cutoff frequency of 4096~Hz and a transform window function of
approximately 1~Hz.  The baseline power is approximately 0.3\% below
the expected value of 2.0 (Leahy et al. 1983) due to an instrumental deadtime 
of approximately 10~$\mu$s. No additional cuts were made on the PCA
energy channels.

In order to calculate count rates and RMS fractions we must correct
for the time-varying background, which is 100 to 150 c/s compared to a
total source count rate of 400 to 700~c/s. We first note that in earth
occultation, the background rejection channels returned as PCA data products
(e.g.  Very Large Event triggers, `VLE', or 6-fold anti-coincidence
triggers) are well correlated with the good count rate.  We calculate
the linear fit of VLE rate versus the `good event' count rate for
Standard Mode 2 data from all the data in occultation for a given day
in one of the PCUs.  This is done using no channel cuts and matching
the number of active PCUs to establish the calibration (in some
intervals only four of the five PCUs were active).  The Standard Mode
count rate is about 2 c/s higher than the Event Mode rate since 5
fewer high energy channels are used. We correct for this small
difference. The final result is a background rate estimate for the
Event Mode data in 16 second intervals, which we use to calculate the
source count rate.  The errors introduced by statistical uncertainty
in the background calibration are small.

\section{Results}

A typical power spectrum from a 2800~s interval is shown in Figure~1.
The novel features of the power spectra are the peaks above
500~Hz. Figure~2 displays the high frequency portions of power spectra
from various intervals.  Two highly significant peaks are
simultaneously present in the power spectra at high count rates (above
approximately 400~c/s).  At lower rates, a single high frequency peak
is visible.  We parameterize these peaks with Lorentzians, which
provide good fits in all cases with typical $\chi_\nu^2$ of
approximately 1.

The frequency centroids of the QPOs are strongly dependent on the
count rate, $R$, as shown in Figure 3. We identify a high and low
frequency QPO whose motions in the $R$ vs $\nu$ plane are clearly
distinct.  The $R$ vs $\nu$ relation of the QPOs from the April
observation (Figure 3) can be fit be power laws with slopes,
$d\log{\nu}/d\log{R}$, of $0.79\pm 0.09$ (high frequency peak) and
$1.17\pm 0.10$ (low frequency peak).  Linear fits are not
statistically preferable.  In the August observation the QPOs occupy a
different place in the $R$ vs $\nu$ diagram.  The count rates are
smaller for a given frequency and deviate from a power law relation at
small $R$.  Above approximately 550~c/s, the correlation of $R$ with
$\nu$ in the August data can be fit by power laws with exponents
consistent with the April fits.

The fractional RMS amplitude of the high frequency QPO falls from
approximately 12\% at 400 c/s to 6\% at 600 to 700 c/s. The RMS
amplitude of the low frequency peak varies between 3\% and 9\%. The
measured Q values ($\nu$/FWHM) range from 5 to 20 for the higher
frequency peak and 10 to 40 for the low frequency QPO.  However, the
rate variations in each interval contribute significantly to these
widths. Using the $R$--$\nu$ correlation to account for this
contribution, we estimate that the intrinsic Q of both QPOs is in the
range 10--20. The Q values increase somewhat as the count rate rises,
while the FWHM of the QPOs decreases.

The difference between the frequency centroids of the two QPOs is
remarkably constant (Figure 4).  The frequency difference from the
April data is $325\pm 5$ Hz.  The frequency difference in the August
observation, $321\pm 6$ Hz, is consistent with that in April even
though the QPOs clearly occupy a different region of the $R$ vs $\nu$
diagram.  Taking all of the data together yields a mean frequency
difference of $323\pm 4$ Hz.

An additional peak is detected at $328\pm 2$~Hz during the interval
beginning UTC 4/24/96 19:47:27 (Figure 2).  This is detected at a
significance greater than 3$\sigma$ and is the only such feature in
the power spectra from 200~Hz to 4000~Hz other than the QPOs discussed
above.  During this interval the two other QPOs are at $549\pm 7$~Hz
and $858\pm 19$~Hz.  The $328 \pm 2$~Hz peak is significantly narrower
(FWHM $\sim 12$~Hz) than the higher frequency peaks. The frequency of
the third peak is consistent with the difference in frequency of the
549 and 858 Hz peaks.

\section{Discussion}

The detection of a constant frequency difference for the two high
frequency QPOs in observations separated by three months clearly
indicates that there is a clock in this system which is stable on at
least this time scale.  The most likely candidate is the neutron star
with an inferred spin period of $3.10\pm 0.04$~msec.  The narrow
feature at $328\pm 2$~Hz may be a direct detection of the neutron star
spin period.  The lack of coherence may be due to reprocessing of the
radiation (Strohmayer et al.  1996).

The leading theory of the origin of low magnetic field millisecond
radio pulsars has long been spin-up by accretion from a companion star
(Alpar et al.  1982).  LMXBs containing low magnetic field neutron
stars are then the progenitors of millisecond radio pulsars and should
contain fast pulsars.  Our detection of a stable 3.1~ms period in
4U~0614+091, together with the detection (Strohmayer et al.  1996) of
a 2.8~ms period in 4U~1728-34 (GX~354-0), provide the missing link in
this evolutionary scenario: direct evidence for millisecond pulsars in
LMXBs.

A detailed discussion of mechanisms of QPO generation are beyond the
scope of this paper.  However, we note that the kinematics of a
magnetospheric beat-frequency model (Alpar and Shaham 1985) gives an
adequate account of the frequencies observed.  In such a model there
are three relevant frequencies in the system: the frequency of
Keplerian orbits at the inner edge of the accretion disk $\nu_K$, the
spin frequency of the neutron star $\nu_S$, and the difference between
these frequencies - the `beat' frequency $\nu_B=\nu_K-\nu_S$.  We
identify the higher frequency peak (Figure 2) as $\nu_K$ and the lower
frequency peak as $\nu_B$.  The frequency of each signal varies as
result of a changing mass accretion rate which alters the accretion
disk geometry.  This model predicts that the frequency difference,
$\nu_K-\nu_B$, is constant and is equal to the spin of the neutron
star.

In the beat-frequency model the inner edge of the accretion disk is
taken to be the (accretion rate dependent) magnetospheric radius
(Alpar and Shaham 1985, Lamb et al. 1985, Ghosh and Lamb 1992).
However, the simplest version of this model predicts a relation
between the Keplerian frequency and the count rate, $\nu _K \propto
R^{\alpha}$, with $\alpha = 3/7$, while our observations show a
significantly steeper power law and a deviation from the power law at
low count rates.

Recently, Miller et al.  (1996) have considered a model in which QPOs
are generated at the sonic point in the accretion disk flow, and a
radiation feedback mechanism drives the beat-frequency signal.  This
model seems adequate to explain the large coherence, large RMS
amplitudes, and steep $R$ vs $\nu$ relation of the high frequency
QPOs.  In this model, the higher frequency QPO is also identified with
a Keplerian orbital frequency.

Two high frequency QPOs with a varying frequency difference have been
observed from the Z-source Sco X-1 (van der Klis et al.  1996a).  The
variation of the frequency difference in this source is in marked
contrast to the constancy of the frequency difference over a broad
span in time and luminosity in 4U~0614+091.  The RMS amplitudes of the
QPOs in Sco X-1 are significantly smaller (approximately 1\%) than
those in 4U~0614+091, and Sco X-1 has a much higher luminosity (close
to Eddington) and a higher magnetic field.  These differences suggest
different origins of the QPOs in Sco X-1 and 4U~0614+091.  We note
that the photon bubble oscillation model (e.g.  Klein et al.  1996)
being applied to high luminosity and high field sources such as Sco
X-1 does not have a natural means to produce a QPO frequency
difference which is constant on time scale of months as observed in
4U~0614+091.

The behavior of the QPOs in 4U~0614+091 are apparently most similar to
those in the atoll source 4U~1728-34 (Strohmayer et al.  1996).  Two
QPOs are observed from 4U~1728-34 at approximately the same
frequencies scaling with count rate over a wide dynamic range.
However, the QPOs in 4U~1728-34 appear at a higher count rate and may
have a steeper $R$ vs $\nu$ correlation.

If the highest frequency QPO is identified as a Keplerian orbital
frequency, then our measurement of a frequency centroid of $1144.7 \pm
9.6$~Hz for the 1800~s interval beginning 8/6/96 20:52:01 UTC can be
used to constrain the mass and radius of the neutron star in
4U~0614+091. In a Schwarzschild spacetime (an adequate approximation
given the 3.1~ms period of the neutron star), no stable orbits exist
within a radius of $6GM/c^{2}$.  Observation of an orbital frequency
$\nu_{K}$ then places an upper limit on the mass of the compact object
of $M = c^{3}(12 \sqrt{6} \pi G \nu_{K})^{-1} = 2.2 M_{\odot}
(\nu_{K}/1000\, {\rm Hz})^{-1}$.  Therefore, the mass of the neutron
star in 4U~0614+091 must be less than $1.9 M_{\odot}$.  The radius of
a circular orbit is simply $r = (GM/4\pi^{2} \nu_{K}^{2})^{1/3}$,
which implies an upper limit on the neutron star radius of 17~km for a
mass of $1.9 M_{\odot}$.  We note that disruption of the accretion
disk flow at the marginally stable orbit (Paczynski 1987) would cause
the frequency versus count rate relation (e.g.  Figure~3) to flatten
above a critical frequency (Miller et al. 1996).  It is interesting to
note that the maximum frequencies observed in the sources 4U~0614+091,
4U~1636-536 (van der Klis et al. 1996b), 4U~1734-28 are all comparable
and would imply a neutron star masses of $1.9 M_{\odot}$.  Additional
observations of these sources, particularly in high luminosity states,
may provide strong constraints on the properties of neutron stars.

\acknowledgments

We thank the staff of the XTE Guest Observer Facility, particularly
Alan Smale and Gail Rohrbach, for assitance with the XTE data.  We
thank J. Swank, F.K.  Lamb, J. Halpern, K. Chen, M. Ruderman, and
C. Miller for useful discussions.


\clearpage

\begin{figure*} \figurenum{1} \epsscale{2.0} \plotone{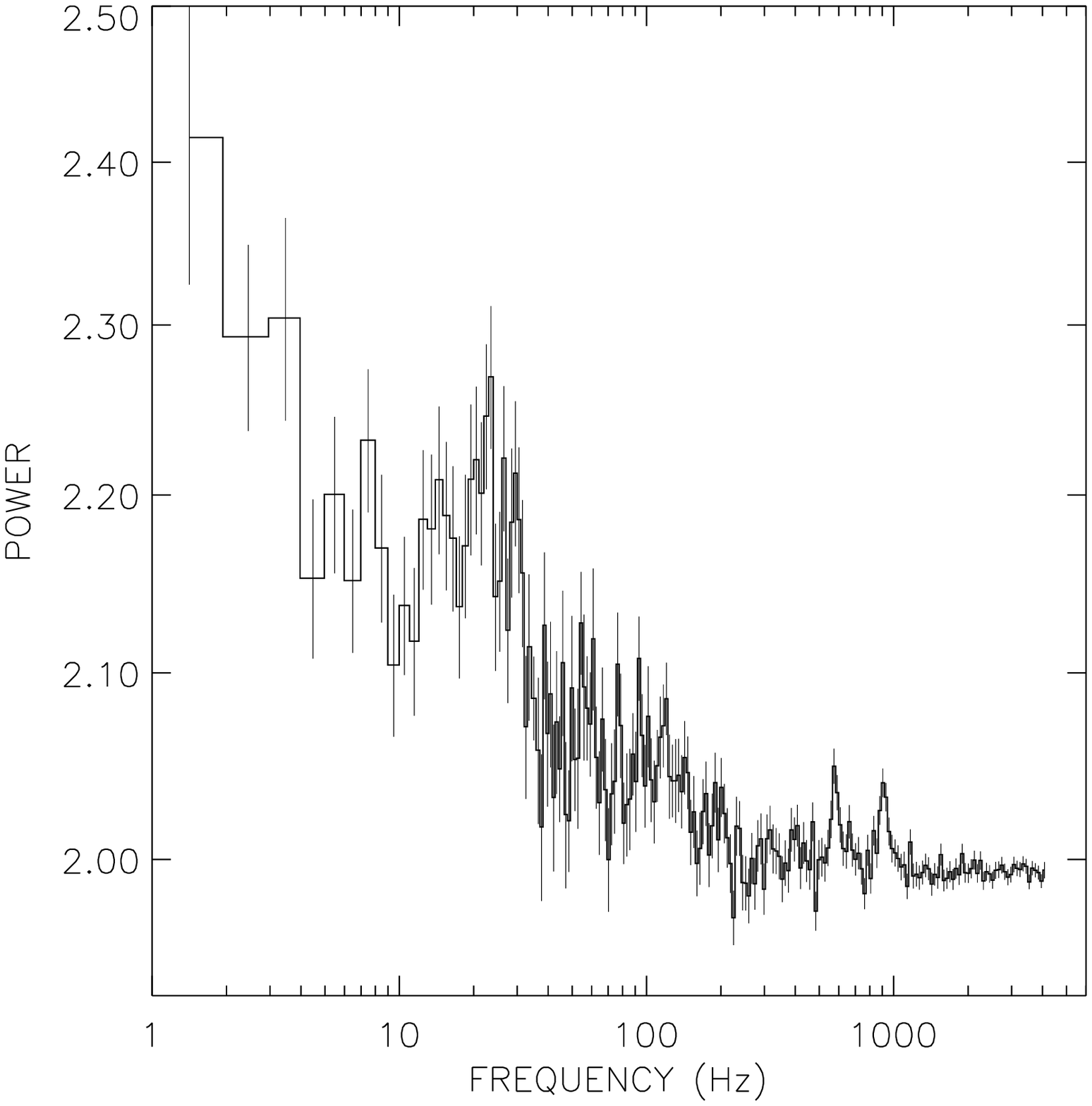}
\caption{Power density spectra of 4U~0614+091 for the 2783~s interval
beginning UTC 4/25/96 4:58:23.  Normalization of Leahy et al.  (1983)
has been used.} \label{fig:pds_full} \end{figure*}

\begin{figure*} \figurenum{2} \epsscale{2.0} \plotone{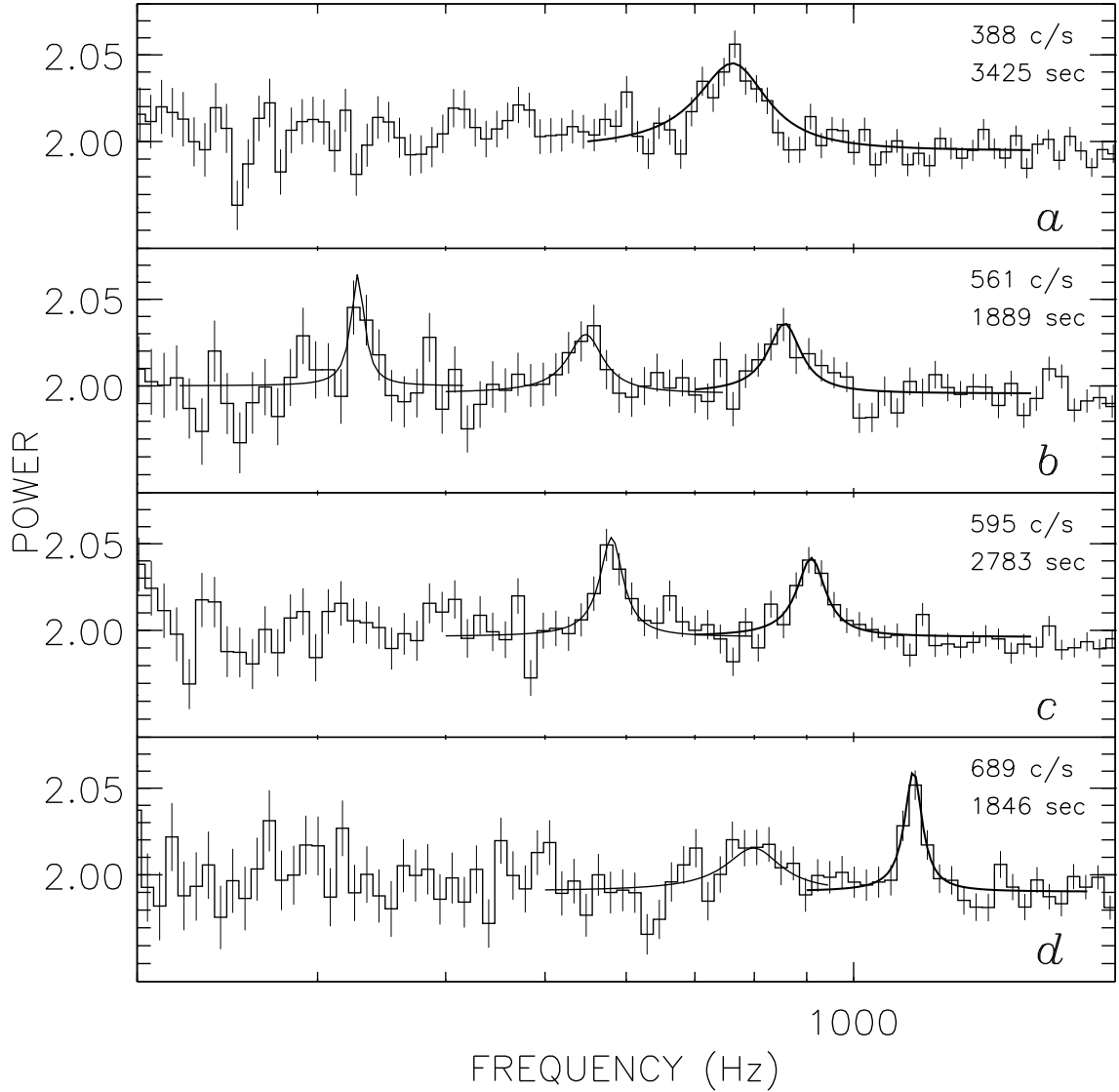}
\caption{Power density spectra of 4U~0614+091 for five intervals
beginning UTC 4/25/96 0:10:23 (a), 4/24 19:47:27 (b), 4/25 4:58:23
(c), and 8/6 20:52:01 (d). The observation time for each spectrum and
the total count rates are given. Fits are shown to the high frequency
(thick line) and lower frequency (thin line) QPOs.}
\label{fig:pds_qpos} \end{figure*}

\begin{figure*} \figurenum{3} \epsscale{2.0} \plotone{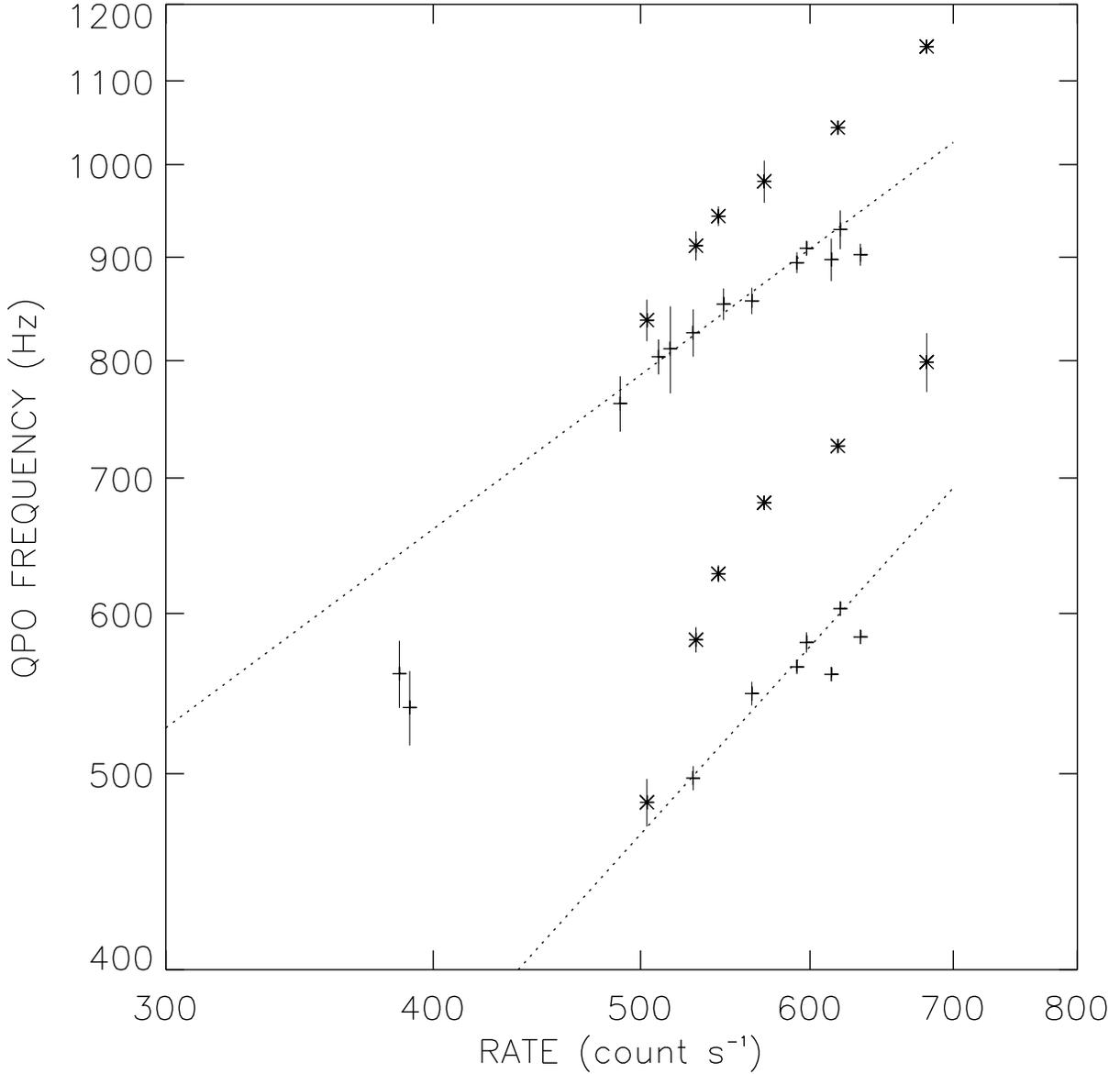}
\caption{QPO centroid frequency versus the total PCA count rate.  The
pluses are April data, and the asterisks are August data.  Power law
relations are fit to the QPO detections between UTC 4/24/96 16:35:27
and 4/25/96 5:44:46.}
\label{fig:freq_rate} \end{figure*}

\begin{figure*} \figurenum{4} \epsscale{2.0} \plotone{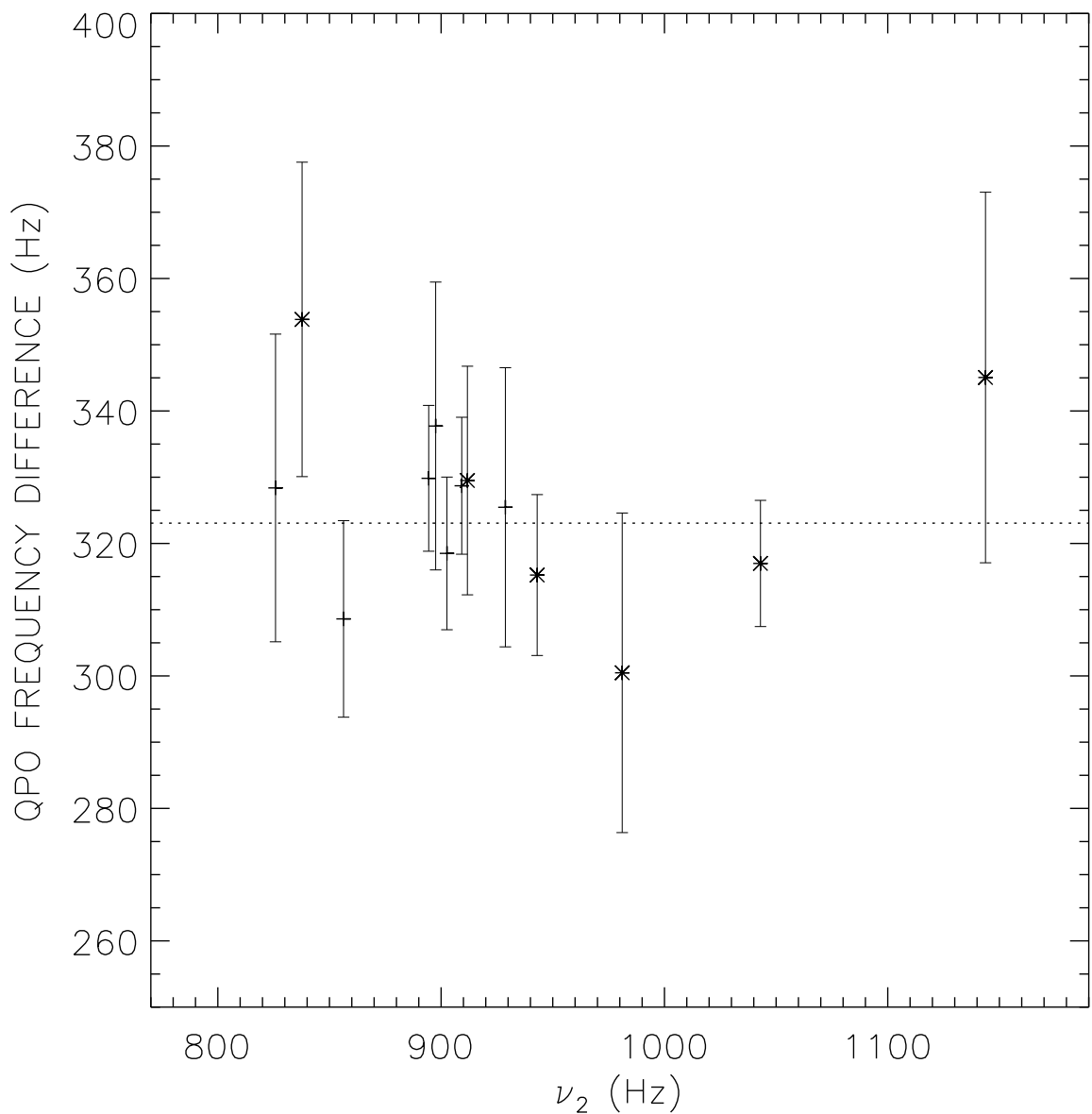}
\caption{Frequency difference of the two simultaneously detected QPOs
vs the frequency centroid of the higher frequency QPO.  The pluses are
April data, and the asterisks are August data.  The mean frequency
difference (dotted line) from all the data is $323\pm 4$ Hz.}
\label{fig:freqdiff} \end{figure*}

\end{document}